\documentclass[useAMS,usenatbib]{mn2e}
\usepackage{graphicx}

\def\simlt{\lower.5ex\hbox{\simlt}}
\def\gtsima{$\; \buildrel > \over \sim \;$}
\def\simgt{\lower.5ex\hbox{\gtsima}}

\title{Red Clump stars in the Bo\"otes~III stellar system}
\author[M. Correnti et al.]
       {M. Correnti $^1$, M. Bellazzini$^2$ and F.R. Ferraro$^1$\\
        $^1$Dip. di Astronomia - Universit\`a di Bologna, Via Ranzani 1, 40127
	Bologna, Italy.\\
        $^2$INAF - Osservatorio Astronomico di Bologna, Via Ranzani 1, 40127
	Bologna, Italy.}
\date{Accepted ...
      Received ...
      in original form ...}

\pagerange{\pageref{firstpage}--\pageref{lastpage}}
\pubyear{2009}

\begin{document}

\maketitle

\label{firstpage}

\begin{abstract}
We report on the detection of a population of Red Clump (RC) stars probably
associated with the recently discovered stellar system Bo\"otes~III. The RC is
identified as a $3\sigma$ peak in the  Luminosity Function (LF) of
colour-selected stars extracted from the SDSS database. The peak is consistently
detected in the $g,r,i$ and $z$ LFs at the expected luminosity   of a typical RC
at the distance of Bo\"otes~III. Moreover the stars around the LF peak show a
maximum of surface density nearly coincident with the reported center of the
system. Assuming that the detected feature is the genuine RC of Bo\"otes~III, we
find that the system has the HB morphology typical of old and metal-poor dwarf
spheroidals, it has an integrated magnitude $M_V\simeq -5.8 \pm 0.5$ and an
ellipticity $\epsilon \sim 0.5$, quite typical of the recently identified new
class of very faint dwarf  galaxies.  
\end{abstract}


\section{Introduction}

The advent of large modern surveys, like the 2 Micron All Sky Survey 
\citep[2MASS,][]{2mass} and the Sloan Digital Sky Survey \citep[SDSS,][and
references therein]{dr6}, has greatly increased our ability to detect
stellar systems and/or structures of  very low surface brightness in the halo
and the disc(s) of the Milky Way (MW). Large scale sub-structures have been
traced over huge portions of  the sky \citep{heidi,yan,maj,cma,juric}, but
also  feeble tidal streams have been found around disrupting globular clusters
\citep[see, for example][]{connie,GJ} or lacking an evident progenitor
\citep{FoS,boo3}. Moreover, a completely new class of very faint  dwarf
galaxies \citep[$0\la M_V\la -8.5$; ][hereafter MJR08]{dwarfs} have been
discovered  \citep[see][and references therein]{fivegal,CVn,UMaII,mike,liu}.
These substructures are generally interpreted as the relics of the
process of hierarchical assembly of the MW, as predicted by galaxy  formation
models within the current cosmological scenario  \citep[see][and references
therein]{bull,madau}.

In \citet[][hereafter B06]{grad} we used SDSS data to study the stellar content
of the largest tidal stream in the halo of the MW, the one produced by the
disruption of the Sagittarius dwarf spheroidal (Sgr dSph) galaxy 
\citep{siba,maj,FoS}. In particular, we showed that it is possible to detect
the Red Clump (RC) of core-He-burning stars associated with a
given sub-structure as a peak in the Luminosity Function (LF) of sub-samples of
stars selected in a relatively narrow colour range including the RC. The
RC peak of the sub-structure can be disentangled from
the fore/background contaminating population of the MW by subtracting the
underlying LF, that is, in general, quite smooth and smoothly varying with
position in the sky. In B06 we used this technique to compare the Horizontal 
Branch (HB) morphology in the stream and  in the main body of Sgr.
However, it was pointed out that the most natural and direct application 
would be the determination of accurate distances from the magnitude
of the detected RC peaks, as the RC is well known and widely used as a standard
candle  \citep[see][and references therein]{pacz,stagar,gs,babu,mrc}. For
intermediate-old age populations, the luminosity of the RC peak shows
relatively modest variations as a function of age and metallicity, in
particular when measured in the reddest optical passbands  \citep[as Cousins'
I, see][]{gs}. We have used theoretical isochrones from the \citet{leo}
set to verify that this is true also for the $r,i,z$ SDSS passbands. We found
that the magnitude of the RC peak varies by $\simeq 0.55$ mag, $\simeq 0.43$
mag, and $\simeq 0.34$ mag, in $r$, $i$ and $z$, respectively, for ages between
4 and 12.5 Gyr over the metallicity range $-1.8\le [M/H]\le 0.0$. 

We have started an extensive search and analysis of the RC peak along the
portion of Sgr stream that is enclosed in the SDSS, a study that is now almost
completed \citep{mat2}. During the analysis, while looking at the LF of a
control field outside the Sgr stream, we noticed a relatively weak but well
defined peak where a smooth LF was expected instead. Following up this finding,
we realized that the considered field enclosed a faint stellar system that was
very recently discovered with a different technique by \citet[][hereafter
G09]{boo3}, i.e. the possible dwarf galaxy (or relic dwarf galaxy) Bo\"otes~III.
G09 detected Boo~III as a weak overdensity centered at (RA,
Dec)=($209.3\degr$,$26.8\degr$) and extending for $\sim 1.5\degr$ from the
center (but most of the signal is detected within $R\sim 0.8\degr$),  at a mean
heliocentric distance $D\sim 46$ kpc. The system has been detected with the
{\em matched filter} technique \citep{connie} that allows the maximization of
the signal from the filtered population over the strong  Galactic
fore/background. The filter adopted by G09 was mostly  sensitive to the blue
side of the Main Sequence (MS) and Turn Off (TO) region of a typical old and
metal-poor population. G09 concluded that ``the galaxy is revealed almost
entirely by subgiant and TO stars...". No significant correlation of the new
system  with structures in the interstellar extinction maps  \citep[][hereafter
SFD98]{sfd} was found, and indeed the reddening is quite low in all the (large)
area explored by G09.  The cluster of galaxies Abell 1824, having a radius of $R
\simeq 13\arcmin$,  was noted to lie within $3\arcmin$ from the object, but G09
showed that the density map obtained from the objects classified as galaxies in
the SDSS does not present any obvious correlation with Boo~III.
The field-subtracted CMD (G09 Fig.~12) shows upper MS, TO and SGB features
that are reasonably bracketed by the ridge lines of the globular clusters M13
and M15, having [Fe/H]$=-1.54$ and [Fe/H]$=-2.26$, respectively.  The RGB
and (possible) RC remain buried in the subtraction noise, that dominates the
plot for $r\la 20.5$.  However the inspection of the un-subtracted CMD allowed
G09 to identify a population of likely BHB stars (at $(g-i)_0<0.0$, where the
contamination is very low) probably associated with Boo~III. Matching the
observed BHB with empirical templates, he found $(m-M)_0=18.35$. 
Boo~III is
very sparse and appears significantly elongated in the E-W  direction. G09
finds it immersed into a very faint stellar stream, dubbed {\em Styx}. Since he
found that the populations in the two substructures appear similar he concluded
that Boo~III is (very likely) a disrupting dwarf galaxy  physically associated
with  the stream and that it is its most  probable progenitor.
In this letter we report on the detection of Bo\"otes~III (Boo~III)  from RC
stars and on the new insight on the stellar content of this system obtained
from this population.
 
\section{The Red Clump of Bo\"otes~III}
To trace the RC population of Boo~III, that was not detected by G09, 
we will adopt the technique developed by B06. The technique used by G09 
\citep[and many others, see][]{connie,heidi, yan,FoS,fivegal} is optimized for
the detection of new structures, as it maximizes the contrast over the
background. It uses mainly MS stars - the most abundant stars in any
stellar system - thus reaching very low surface brightness levels and allowing
the efficient surveying of large areas of the sky. On the other hand, our
approach is best suited for a more accurate localization of detected
structures. The method collapses all the information from colour-selected
candidate RC stars into a mono-dimensional histogram, where any statistically
significant peak provides a direct estimate of the distance. The RC stars are
much less numerous than MS stars but, in general, are tightly {\em clumped}, 
both in colour and in magnitude, when placed at the same distance.
In the following we will consider the SDSS-DR6 photometry of objects classified
as stars \citep[extracted
from the SDSS CasJobs query system,][]{dr6} from  two fields: (a) F1, a circular
field centered on the center of Boo~III (as determined by G09) and with
radius $R=0.8\degr$, (b) CF, a square  $8\degr \times 8\degr$ Control Field
with the same center as F1 but with the inner $R\le 2\degr$ circle removed. To
average out statistic noise in the fore/background population, CF has much
lager area than F1 (by a factor of  $\sim 20$). As the two globular clusters
NGC5466 and M3 were included in CF, we excised from it two circular areas
centered on the clusters and having radius $R=1\degr$, i.e. much larger than
their limiting radii \citep[][H96, hereafter]{h96}.
All the stars are
corrected for reddening using the SFD98 maps. The average and standard deviation
 in 
E(B-V) is $0.019\pm 0.003$ in F1 and  $0.015\pm 0.003$ in CF. In the following
all the reported colours and magnitudes  must be intended as corrected for
reddening.

\begin{figure*}
\includegraphics[width=118mm]{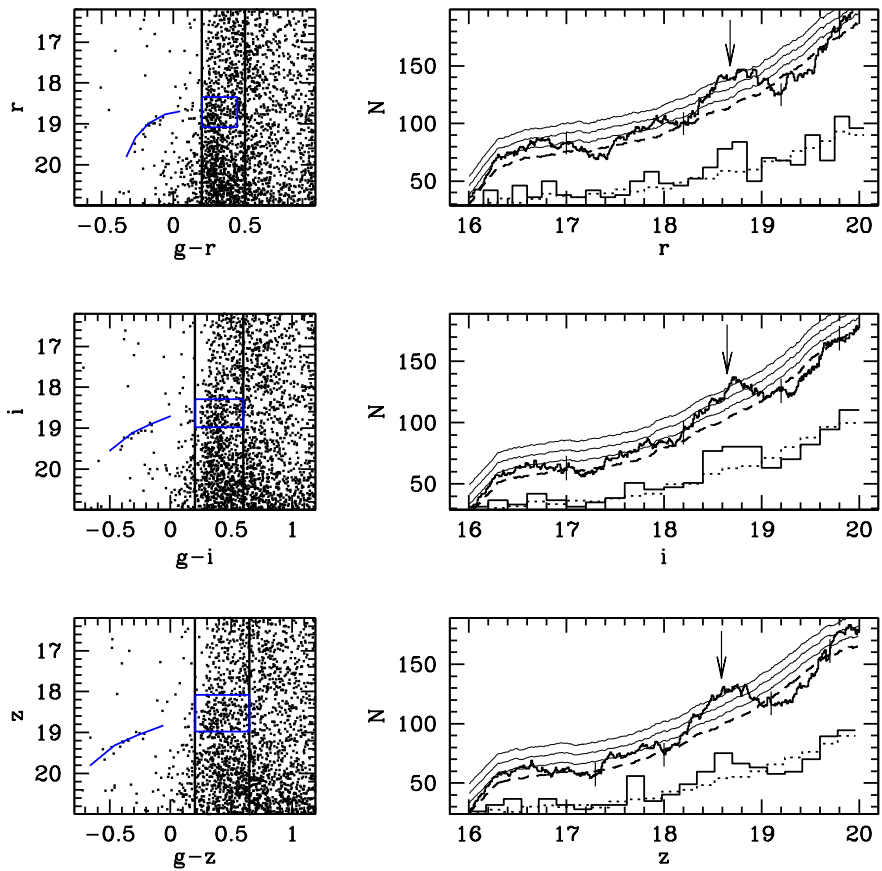}
  \caption{Left panels: CMDs of F1 in various combinations of magnitudes and
  colours. The vertical lines enclose the adopted colour-selection windows. The
  boxes and the ridge-lines are the templates for Draco RC and BHB,
  respectively, as derived in Fig.~\ref{dra}, below, corrected for the distance
  of Boo~III derived here. 
  Right panels: de-reddened running-histogram LFs of colour-selected RC candidates 
  for F1 (continuous line) and CF (dashed line) in $r$, $i$ and $z$. 
  The thin lines marks the 1, 2, and 3$\sigma$ levels above the background, 
  including all the sources of uncertainty. The estimated positions of the 
  peaks are indicated by arrows, the ranges over which we have performed the
  normalization are indicated by thin vertical segments. The LFs represented as
  ordinary histograms are also reported for comparison (F1: thin continuous 
  histogram, CF: thin dotted histogram), after being multiplied by a factor of
  $\simeq 2$ to fit into the limits of the plots.
  }
  \label{lf}
\end{figure*}

The LFs in $r$, $i$, and $z$ of F1 and CF stars, lying in the colour range that
is expected to enclose the RC, are compared in the right panels of
Fig.~\ref{lf}.  The independent colour selections adopted in $(g-r)$, $(g-i)$
and $(g-z)$ are shown in the corresponding CMDs in the left panels of
Fig.~\ref{lf}. The LFs are computed as running histograms having bin width of
$\pm 0.3$ mag and step of $0.01$ mag, as this approach allows the best
determination of the position of peaks \citep[see][and references
therein]{leo2}.    It seems quite clear that the LFs of F1 present additional
structures with respect to CF, for $r,i,z \la 19$. In particular, all the LFs
show a
clear peak around $r,i,z \sim 18.5$ that is hardly compatible with a chance
fluctuation of star counts.  The peak is detected also in the $g$ LF but it is
not shown if Fig.~\ref{lf} for reasons of space. The LFs of the CF have been
normalized to the F1 ones  by minimizing the average difference 
over large  intervals around the main peak, i.e.
$17.0<r,i<18.2$  and $19.2<r,i\le19.8$, $17.3<z<18.0$ and $19.1<z\le19.7$. In
all the considered cases the peaks reach the $3\sigma$ level above the
background, that includes both the Poisson noise and the uncertainty in the
adopted normalization.

We estimated the position of the peaks by fitting Gauss curves to the
residuals of the subtraction between the LFs of F1 and CF. 
We obtain $g=19.05$, $r=18.68$, 
$i=18.65$, and $z=18.59$, with a typical uncertainty of $\pm 0.05$ mag. 
If interpreted as genuine Red Clump peaks, these would imply a distance very
similar to Boo~III. For instance, transforming the theoretical 
absolute V and I magnitudes of the RC for a Z=0.001, age=10 Gyr population
provided by \citet{gs} into $M_r^{RC}=0.40$ and $M_i^{RC}=0.23$\footnote{Using
transformations derived from the comparison between 713 stars in common between
the SDSS and \citet{stet} photometry of NGC~2419.}, we find
$(m-M)_0=18.28$ and $ 18.42$, to be compared with $(m-M)_0=18.35$ (G09).
The coincidence calls for a common
origin of the MS + SGB + BHB populations found by G09 and the possible RC
detected here. 

When dealing with such distant, ghostly systems, with very low surface 
brightness and luminosity, like Boo~III  \citep[or Boo~II, CV~II, etc., see
MJR08,][]{liu}, we may face the possibility that the proof of their existence
and their characterization can be obtained only by painstaking accumulation of
many - sometime weak, but consistent - clues. While the evidence presented here
is not sufficient in itself to confirm the presence of a stellar system, we
think that it provides strong support to the findings by G09 and it may allow a
further insight on the nature of Boo~III. In the following we will show that
even if the statistical significance of the detected peak is not very strong
($3\sigma$), the interpretation of this feature as the  RC population of Boo~III
has an excellent degree of consistency with the  known properties of the stellar
system.

Even if in Fig.~\ref{lf} we adopted the intervals that maximize the signal of
the RC peak, the feature is well detected for a range of colour selections. On
the other hand the colour of the RC is mostly sensitive  to the chemical
composition \citep{gs} and can provide useful indications on the metallicity of
the detected population. To allow a direct comparison with empirical templates
we studied the RC population of the two old clusters displaying an RC in the
SDSS sample by  \citet[][i.e., Pal~4, M71]{an}.  The colour of the putative RC
of Boo~III ($g-r=0.37$) is much bluer than  that of M71 ($g-r=0.62$ and
$[Fe/H]=-0.73$), and even bluer than that of Pal~4 ($g-r=0.48$ and
$[Fe/H]=-1.48$), suggesting that  these stars belongs to a population   more
metal-poor than $[Fe/H]\sim -1.5$, in agreement with the independent results by
G09. 

\begin{figure}
\includegraphics[width=80mm]{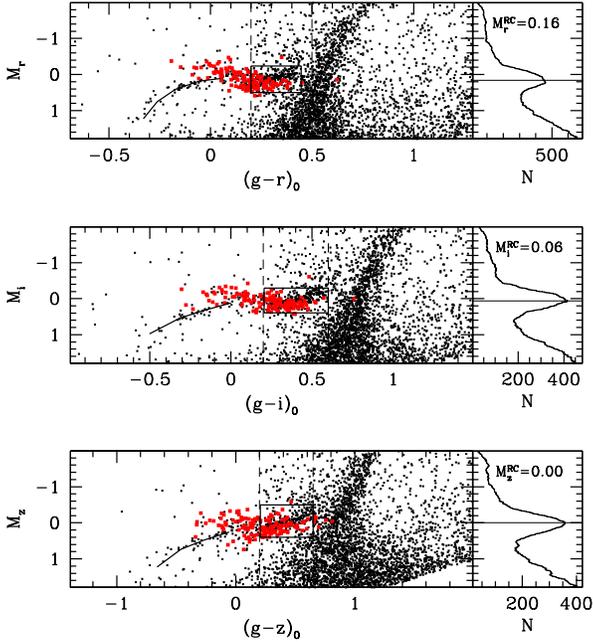}
 \caption{CMDs of the innermost $R=15\arcmin$ of the Draco dSph from photometry
 extracted from the SDSS-DR6. The heavier (red) points are RR Lyrae
 counter-identified from the set by \citet{bona}.  The rectangles that
 approximately enclose the RC in the various CMDs, and the BHB ridge lines have
 been drawn by eye. The thin dashed lines show the  same colour windows adopted
 in Fig.~\ref{lf} to select candidate RC stars.  The same is also adopted for
 the LFs shown in the right panels, with the  same step and bin width as in
 Fig.~\ref{lf}.
 }
 \label{dra}
\end{figure}

If Boo~III is a new stellar system and the LF peak detected here is indeed its
RC population, we are facing an old metal-poor system that displays a complex
(bi-modal?) HB morphology,  having  both a BHB and a RC. This is very
reminiscent of classical dwarf spheroidal satellites of the MW dominated by old
and metal-poor populations like, for example,  Sculptor, Sextans, Leo~II,
Tucana, Ursa Minor and Draco \citep[see][and references therein]{harb}.   Since
Draco is included in the SDSS it may provide the ideal template for a
simultaneous fit of the RC and BHB of Boo~III. In Fig.~\ref{dra} we present a
determination of the $r$, $i$, and $z$ absolute magnitudes of the RC in Draco,
obtained from the peak of colour-selected LFs, as above, from SDSS-DR6
photometry. The adopted reddening and distance are from \citet[][hereafter
B02]{dra},  the RR Lyrae are identified from the list by \citet{bona}. We have
also plotted rectangles that approximately comprise the RC  in the various CMDs,
as well as ridge lines fitting the mean loci of the BHB. With these tools in
hand we can attempt a simultaneous fit of the RC and BHB of Boo~III. In case of
success this will lend strong support to the association of the two features
and, in turn, to the interpretation of Boo~III as a dwarf galaxy (or relic of).
Coupling the apparent magnitudes of Boo~III RC from Fig.~\ref{lf}  with the
absolute magnitudes of Draco RC from Fig.~\ref{dra} we obtain $(m-M)_0=18.46$,
$18.59$ and $18.59$, from $r$, $i$, and $z$, respectively. Averaging these
value we obtain a final estimate of  $(m-M)_0=18.58 \pm 0.05$ (internal) $\pm
0.14$(external), in good agreement  with G09, once taken into account that his
estimate is based on the H96  distance scale that is $\simeq 0.2$ shorter than
the scale adopted in B06 \citep{F99}. Accounting for this systematic the two
estimates differ by only 0.03 mag. Shifting the Draco RC+BHB templates by
$(m-M)_0=18.58$ in the Boo~III CMDs we achieve the matches of the BHB shown in
the left panels of Fig.~\ref{lf}, that can be considered as very satisfactory.
Hence the BHB and the putative RC of  Boo~III can be simultaneously fitted by
the respective Draco templates. Fig.~\ref{dra} also indicates that (a) our
colour-selected RC samples may be contaminated by type ab RR Lyrae observed at
random phase, that may partially blur the actual RC signal, and (b) the weak
features in the F1 LFs, brighter than the RC peak, may be due to Asymptotic
Giant Branch stars falling into the selection window.

\begin{figure}
\includegraphics[width=70mm]{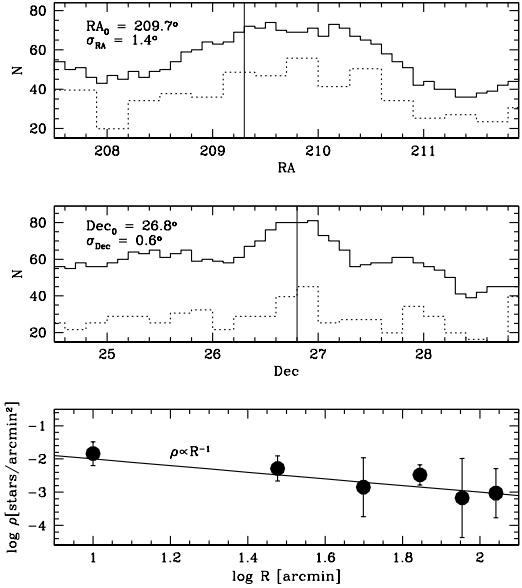}
 \caption{Spatial distribution of stars around the RC peak (colour selections
 as in Fig.~\ref{lf} and $18.5\le r\le 18.8$). Upper panel: distribution in RA
 within a $2\deg$ wide strip enclosing Boo~III (Dec=$26.8\degr \pm 1\degr$). 
 Middle panel: 
 distribution in Dec within a $2\deg$ wide strip enclosing Boo~III 
 (RA=$209.3\degr \pm 1\degr$). The horizontal scale is the same in the two
 panels. The adopted bin width and step of the running histograms (continuous
 lines) are $\pm 0.4\degr$ and $0.1\degr$, respectively. The ordinary hitograms
 (dotted lines, rescaled by a factor of $\times 1.8$) are also reported, as a
 reference.
 The vertical thin lines marks the position of the center of Boo~III 
 estimated by G09. The mean and $\sigma$ of the Gauss curves best fitting our
 distributions are reported in the upper-left corners. 
 Lower panel: the background-subtracted azymutally-averaged density profile is
 compared with a $R^{-1}$ profile.}
 \label{spaz}
\end{figure}

In the upper panels of  Fig.~\ref{spaz} we show the running histograms of RA and
Dec for stars in the RC peak. The two distributions have clear maxima nearly
coinciding with the position of Boo~III estimated by G09.  Moreover the Gauss
curve that best fits the peak in the RA direction  has a $\sigma$ $\sim 2$
times larger than its counterpart in Dec, nicely confirming the strong E-W
elongation found by G09. Finally,  the plot in the lower panel of
Fig.~\ref{spaz} shows that the surface density  profile of RC peak stars
displays a radial decline from the center of Boo~III, with a slope fully
consistent with what found by G09.

\section{Discussion}

We have detected a  peak in the LFs  of
colour-selected candidate RC stars in a circular $R\le 0.8\degr$ field 
centered on the newly proposed dwarf galaxy Boo~III (G09). 
Even if the statistical significancy of the peak is not particularly strong ($3
\sigma$),
the detection (a)
has been consistently obtained in the $g,r,i$, and $z$ LFs, (b) in all cases it
is located at the expected luminosity of Boo~III RC, (c) a simultaneous fit of
the putative RC and BHB of Boo~III has been obtained with RC+BHB templates of
Draco dSph, (d) the stars located in the RC peak have a maximum in surface
density at the same position in the sky as Boo~III, and (e) their radial
surface-density profile is  fully consistent with that found by G09 for MS
stars of Boo~III. Hence, it is extremely likely that the  detected signal is
due to a population that is physically associated with the newly discovered
system. This results lend support to the G09  interpretation of Boo~III as a
new stellar system, possibly a disrupting  dwarf spheroidal galaxy, even if a
final verdict must wait for spectroscopy of a sizable sample of member stars. 
If we accept the hypothesis that Bo\"otes~III is a genuine stellar system, from
the present analysis we can draw the following conclusions:
\begin{enumerate}

\item The system has a composite HB morphology very typical of dSph galaxies
dominated by old and metal-poor populations (but having a spread in age and
metallicity, like Scu, Sex, UMi, Dra). 
All of these galaxies contains also RR Lyrae variables. By analogy some
RR Ly could be present also in Boo~III. Re-scaling the population of Draco to 
the total luminosity of Boo~III, $\sim 10$ RR Ly are expected.

\item The average distance modulus from the RC is $(m-M)_0=18.58 \pm 0.15$ 
corresponding to $D\simeq 52.0 \pm 3.6$ kpc. The difference with respect to G09
is completely accounted for by the different distance scales adopted. The
detected signal is too weak to look for the difference in distance between the
Eastern and Western lobes of the galaxy claimed by G09. For the same reason we
were unable to find any detectable RC signal from the much fainter {\em Styx} 
stream that was claimed to be associated with the galaxy (G09).  
\footnote{On the other hand,
the LFs of  sources, colour-selected as in Fig.~\ref{lf} and classified 
as ``galaxies'' in the SDSS do not show any peak over the whole considered range
and lie much below than the level reached by stellar LFs
around $r,i,z\simeq 18.5$ (by a factor of $\sim 2$), thus confirming the
conclusion by G09 that the Boo~III overdensity is not associated with the A1824
cluster of galaxies.}

\item The technique adopted here allows us to count the RC stars (B06). Adding
also the BHB we obtain a rough estimate of the number of core-He-burning stars 
($N_{Hb}\simeq 38$) that can
be directly converted into a distance-independent estimate  of the absolute
integrated V magnitude by means of the Evolutionary Flux Theorem \citep{rbuz},
as done in \citet{mrc}. Using Eq.~1 of \citet{alvio}, adopting the parameters
appropriate for a population with [M/H]$=-1.35$ and age=10 Gyr from
\citet[][]{claudia},
and $t_{Hb}=9\times 10^7$ yr from the BASTI database \citep[canonical Z=0.01
models of age=8.2 Gyr;][]{basti}, 
we get  $M_V\simeq -5.8 \pm 0.5$, 
very weakly depending on the assumed age and metallicity. This value is
straight in the middle of the  range covered by the new class of 
faint dwarf galaxies
and in agreement with their distribution in the half-light radius $r_h$ vs.
$M_V$ plane, for any reasonable choice of $r_h$ in the range between 0.1 kpc
and 0.7 kpc, corresponding to $\sim 0.1\degr$ and $0.8\degr$ (see MJR08, and
references therein). From $N_{Hb}$ we also estimated an average surface
brightness within $R\le 0.8\degr$ of $\mu_V\simeq 31.3 \pm 0.3$ mag/arcsec$^2$.
Finally, estimating the ellipticity as $\epsilon =
1-\sigma_{Dec}/\sigma_{RA}$  (thanks to the E-W orientation of the major axis, 
see G08 and Fig.~\ref{spaz}) we find $\epsilon \sim 0.5$,
also in agreement with the other newly discovered  faint dwarfs (MJR08). 

\end{enumerate}

\section*{Acknowledgments}

This research is supported by INAF through the PRIN-2007 grant CRA 1.06.10.04.
We are grateful to R. Ibata for a critical reading of the original manuscript.

\label{lastpage}

\end{document}